\documentclass[a4paper,11pt]{article}
\usepackage{geometry}\usepackage[utf8]{inputenc}
\usepackage{amsmath}\usepackage{amssymb}
\usepackage{xcolor}\usepackage{graphicx}

\title{Chern-Simons Theory in SIM(1) Superspace}

\author{
Ji\v{r}\'{i} Voh\'{a}nka$^1$ and Mir Faizal$^2$ \\
$^1$Department of Theoretical Physics and Astrophysics, Masaryk University,\\
Kotl\'{a}\v{r}sk\'{a} 267/2, 611 37 Brno, Czech Republic \\ 
$^2$Department of Physics and Astronomy,   University of Waterloo, \\
Waterloo, Ontario N2L 3G1, Canada 
}
\date{}
\begin{document}

\newcommand{\fsl}[1]{#1\kern-0.50em/}
\newcommand{\fsnabla}{\nabla\kern-0.75em/}
\newcommand{\di}{\text{d}}
\newcommand{\pxu}[1]{\partial_\times{}^{#1}}
\newcommand{\pxd}[1]{\partial_{\times{#1}}}
\newcommand{\pxm}{\partial_{\times-}}
\newcommand{\gp}{\gamma_+}
\newcommand{\gpp}{\gamma_{++}}
\renewcommand{\wp}{w_+}
\newcommand{\fpp}{f_{++}}
\newcommand{\gx}{\gamma_\times}
\newcommand{\gxp}{\gamma_{\times+}}
\newcommand{\gxx}{\gamma_{\times\times}}
\newcommand{\wm}{w_-}
\newcommand{\fmp}{f_{+-}}
\newcommand{\fmm}{f_{--}}
\newcommand{\gm}{\gamma_-}
\newcommand{\gmp}{\gamma_{+-}}
\newcommand{\gmm}{\gamma_{--}}
\newcommand{\pp}{d_+}
\newcommand{\ppp}{\partial_{++}}
\newcommand{\pmp}{\partial_{+-}}
\newcommand{\pmm}{\partial_{--}}
\newcommand{\pxx}{\tfrac{\Box}{\partial_{++}}}
\newcommand{\exl}[2]{\underset{\text{#1}}{#2}}
\newcommand{\km}{\kappa_-}
\newcommand{\kx}{\kappa_\times}

\maketitle
\begin{abstract}
In this paper, we will analyse a three dimensional 
supersymmetric Chern-Simons  theory   in $SIM(1)$ superspace formalism. The breaking of the Lorentz symmetry down to the $SIM(1)$ symmetry, breaks half the supersymmetry of the Lorentz invariant theory. So, the supersymmetry  of the Lorentz invariant Chern-Simons theory with $\mathcal{N} =1$ supersymmetry will break down to   $\mathcal{N} = 1/2$ supersymmetry, when the Lorentz symmetry is broken down to the $SIM(1)$ symmetry. First, we will write the Chern-Simons action using $SIM(1)$ projections of $\mathcal{N} =1$ superfields. However, as the $SIM(1)$ transformations of these   projections are very complicated, we will define $SIM(1)$ superfields which transform simply under $SIM(1)$ transformations. We will then express the Chern-Simons action using  these  $SIM(1)$ superfields.  
 Furthermore, we will analyse the gauge symmetry of this Chern-Simons theory.  
 This is the
first time    that a Chern-Simons theory with    $\mathcal{N} =1/2$ supersymmetry will be constructed 
on a manifold without a boundary. 
\end{abstract}

\section{Introduction}

Chern-Simons theories are topological field theories in which the action is proportional to the integral of the Chern–Simons 3-form
\cite{ch}-\cite{ch00}.   Chern-Simons theories have important condensed matter applications as they are related to the fractional quantum hall effect
\cite{qf}-\cite{fq}. 
In the fractional quantum hall effect 
collective state in which electrons bind magnetic flux lines to make new quasi-particles, the excitations have
a fractional elementary charge.  The
supersymmetric generalization of the fractional quantum hall effect has also been analysed
\cite{sqhe}-\cite{sqhe1}. 
The Chern-Simons theory has  been used  for studding    inflationary cosmology \cite{in00}-\cite{00in}. In fact, the  
 Chern-Simons theory has also been used to balance  potential forces
in a generic mechanism of inflation  
\cite{in}.  
In this model of inflation the field motion
presence of a large Chern-Simons coupling 
is the direct analogue
of the  magnetic drift of a charged particle in a strong magnetic field. 
Thus, this model of inflation no special assumption is needed for   the kinetic energy and the potential energy terms. 
All that is required is  to make the magnetic drift slow enough to generate a long inflationary era.
This is accomplished by making the Chern-Simons interaction sufficiently large.

Chern-Simons theories are essential for  constructing  the action of multiple M2-branes.  
According to the $AdS/CFT$ correspondence, the superconformal field theory    
 dual to the eleven dimensional supergravity  on $AdS_4 \times S^7$ has $\mathcal{N} =8$ supersymmetry.
This is because   
$AdS_4 \times S^7 \sim [SO(2,3)/ SO (1, 3)]\times  [SO(8)/ SO(7)] \subset OSp(8|4)/[SO(1,3) \times SO(7)]$, and this 
  supergroup $OSp(8|4)$ gets  realized as $\mathcal{N} = 8$ supersymmetry of this dual superconformal field theory. 
  Thus, a requirement for the 
 superconformal field theory describing  multiple M2-branes is that it should  have $\mathcal{N} =8$ supersymmetry.
 Furthermore, this theory should have eight gauged valued scalar fields and sixteen physical fermions. 
 This exhausts all the on-shell degrees of freedom and hence the gauge fields of this theory cannot contribute 
 to any on-shell degrees of freedom. In other words, the gauge sector of this theory should be described by a topological field 
 theory. It is possible to demonstrate that a matter-Chern-Simons theory, called the 
 BLG theory,   satisfies all these properties  
 \cite{1}-\cite{5}. The gauge sector of this theory is described by a Chern-Simons theory in which the gauge fields take  
 values in a   Lie $3$-algebra rather than a conventional Lie algebra.
However,  only one finite dimensional example of 
such a  Lie $3$-algebra exists, so,  this theory only describes two M2-branes. 
It is possible to relax the requirement
of manifest $\mathcal{N} =8$ supersymmetry, and use the
ABJM theory to study multiple M2-branes.  The gauge sector of the ABJM theory is described by two regular Chern-Simons theories 
 with   levels $k$ and $-k$ \cite{apjm}-\cite{abjm1}. Even though
 it only has 
manifest
$\mathcal{N} = 6$ supersymmetry, its supersymmetry can be enhanced 
to the full $\mathcal{N} = 8$ supersymmetry by monopole operators  for
$k=1$ or $k=2$ \cite{mp}-\cite{pm}. In fact, the ABJM theory
coincides with the BLG theory for the only known example of 
the Lie $3$-algebra. Apart from the constructing of the gauge sector of multiple M2-branes, 
Chern-Simons theories have also been used for analysing  open strings ending on a D-brane in 
the A-model topological string theory \cite{amodel}. The holomorphic Chern–Simons theory
is also used for analysing the B-model in the string theory \cite{bmodel}.

It may be noted that M2-branes can be coupled to background   fields \cite{flux}-\cite{flux1}. 
It is known that 
a noncommutative deformation of field theories occurs due to a constant background $NS-NS$ B-field 
 \cite{az}-\cite{Ferrara:2000mm}. It is also possible to study  
 D-branes in presence of a
$RR$ background  \cite{Ooguri:2003tt}-\cite{4z}.
In fact, a 
gravity dual of such a field theory has been constructed \cite{Chu:2008qa}.
This gives rise to a non-anticommutative deformation of the field theory,  which in turn breaks half of its  supersymmetry. 
Four dimensional theories with $\mathcal{N} =1/2 $ supersymmetry have been constructed
by using  non-anticommutative  deformations of theories with $\mathcal{N} =1 $ supersymmetry \cite{non}-\cite{non1}.
On the other hand, it is not possible to construct a three dimensional theory with $\mathcal{N} =1/2 $ supersymmetry
using a non-anticommutative deformation of the superspace. This is because 
there are not enough degrees of freedom in the three dimensional $\mathcal{N} = 1 $ superspace to perform such a 
deformation. However, it is possible to  use non-anticommutativity  to break the supersymmetry of a 
three dimensional theory from  $\mathcal{N} = 2 $ supersymmetry to 
  $\mathcal{N} = 1 $ supersymmetry  \cite{nonnon}. 
It may be noted that the boundary  effects can  break the supersymmetry of a three dimensional theory with  
$\mathcal{N} = 1 $ supersymmetry to $\mathcal{N} = 1/2 $ supersymmetry \cite{bou}. 
This happens as the supersymmetric variation of a Lagrangian with $\mathcal{N} = 1 $ supersymmetry is a total derivative. 
In presence of a boundary, this total derivative gives rise to a boundary piece breaking the supersymmetry.
However, it  is possible to add a boundary term to 
the original Lagrangian such that its supersymmetric variation  exactly cancels the boundary piece generated by 
the supersymmetric variation of the original Lagrangian. This way half the supersymmetry of the original theory 
can be preserved. In absence of a boundary the only known way to construct a three dimensional 
 theory with  $\mathcal{N} = 1/2 $ supersymmetry is by breaking the Lorentz group down to the $SIM(1)$ group \cite{sim1}. 
 This is because the breaking of the Lorentz symmetry to the $SIM(1)$ symmetry breaks half the supersymmetry of the  theory 
 \cite{sim1}. In fact, 
  half the supersymmetry of a four dimensional supersymmetric 
theory is also broken, if the Lorentz symmetry is broken down to the $SIM(2)$
symmetry \cite{sups}. 

Various approaches to quantum gravity like 
  discrete spacetime \cite{1q},   spacetime foam models \cite{2q}, spin-networks in loop quantum gravity \cite{4q},
non-commutative geometry \cite{5q}, and Horava-Lifshitz gravity \cite{6q}, predict that the Lorentz symmetry will be broken at the Planck scale. So, 
there are strong theoretical motivations to study  gauge theories in a spacetime where  the Lorentz symmetry is spontaneously 
broken. Thus,    there are strong indications that the Lorentz symmetry  might be only a low energy effective symmetry.  
Even in string theory spontaneous breaking of the Lorentz symmetry  occurs due to an unstable perturbative string vacuum. 
This is because in string field theory a tachyon field has the wrong sign for its mass squared, 
and this causes the perturbative string vacuum to become unstable. If the vacuum  expectation value of the 
tachyon field is infinite, the theory   becomes  ill defined. However, if the vacuum expectation value of the
tachyon field is finite and negative, the coefficient of the quadratic term for the
massless vector field also becomes nonzero and negative. This causes spontaneous
breaking of the Lorentz symmetry to occur \cite{mtheory}. 
It has been demonstrated that  appropriate fluxes  can also break the Lorentz symmetry  in  M-theory \cite{a1}. 
It is also possible to argue for   spontaneous breaking of the Lorentz symmetry in string theory using the 
low energy effective action. Thus, it is possible to analyse 
a gravitational version of the Higgs mechanism for the 
 low energy effective field theory  action obtained from string theory. This gravitational version of the Higgs mechanism 
causes  spontaneous breaking of the Lorentz symmetry  to occur \cite{2a}.

 Motivated by the theoretical developments, which predict the breaking of the Lorentz symmetry in the ultraviolet 
 limit, a model for spacetime geometry has been proposed which only preserves a  subgroups
of the Lorentz group \cite{vsr}. However, the symmetry it preserves is enough to explain the various 
experimental bounds like the constancy of the velocity of light. This theory is called the 
  very special relativity (VSR). In VSR, if the CP symmetry is  also postulated to be a symmetry
  of the theory, then the full Lorentz group is recovered. In this context two subgroups of the
 Lorentz group that have been analysed are called the $SIM(2)$  and $HOM(2)$ groups. 
  The Poincare 
symmetry preserved on a noncommutative Moyal plane with light-like noncommutativity  also gives rise to 
VSR \cite{a1}.  It may be noted that 
  abelian gauge symmetry has been analysed in the context of  VSR
\cite{a}. This work has also been recently generalized to include non-abelian gauge theories \cite{b}. 
Four dimensional  supersymmetric theories with the $SIM(2)$ symmetry have also been analysed  \cite{sup}. A superspace construction 
of such supersymmetric theories has also been performed  \cite{sups}, and the corresponding 
  supergraph rules  for these theories have been derived \cite{supa}. 
The  Yang-Mills theory in $SIM(2)$ superspace formalism  have also been studied 
\cite{supg}. This work has been recently used to motivate the study of three dimensional Yang-Mills-matter theory 
in  $SIM(1)$ superspace formalism \cite{sim1}.  The construction of the   three dimensional Yang-Mills-matter theory 
in $SIM(1)$ superspace formalism was simplified by use of  covariant projections. However, it is not possible to analyse the 
Chern-Simons theory by using covariant projections. So, the construction   of the Chern-Simons theory in 
 $SIM(1)$ superspace requires highly non-trivial calculation, and this is what we aim to do in this paper.

\section{Notation}

The spinor notation, the form of supersymmetry generators as well as the definition of gauge covariant derivatives
will be the same as in \cite{sim1}. 
The supersymmetry generator and the super-derivative for a three dimensional theory with $\mathcal{N} =1$ supersymmetry is 
\begin{align}
 Q_\alpha &= \partial_\alpha - (\gamma^a\theta)_\alpha \partial_a
  = \partial_\alpha + \theta^\beta \partial_{\beta\alpha},
&
 D_\alpha &= \partial_\alpha + (\gamma^a\theta)_\alpha \partial_a
  = \partial_\alpha - \theta^\beta \partial_{\beta\alpha},
\end{align}
such that the anticommutator of spinor derivatives is 
\begin{equation}
 \{ D_\alpha , D_\beta \} = -2 \partial_{\alpha\beta}.
\end{equation}
The gauge covariant derivatives can now be expressed as  
\begin{align}
 \nabla_\alpha &= D_\alpha - i\Gamma_\alpha,
&
 \nabla_{\alpha\beta} &= \partial_{\alpha\beta} - i\Gamma_{\alpha\beta},
\end{align}
and the connections are subject to gauge transformations
\begin{align}
 \Gamma'_\alpha &= e^{iK} \Gamma_\alpha e^{-iK} + ie^{iK} \left( D_\alpha e^{-iK} \right),
&
 \Gamma'_{\alpha\beta} &= e^{iK} \Gamma_{\alpha\beta} e^{-iK} + ie^{iK} \left( \partial_{\alpha\beta} e^{-iK} \right).
\end{align}
where $K$ is a real scalar superfield. The (anti)commutators of gauge covariant derivatives are given by 
\begin{subequations}
\begin{align}
\label{gl:ca}
 \{\nabla_\alpha,\nabla_\beta\} &= -2\nabla_{\alpha\beta},
\\
\label{gl:cb}
 [\nabla_\alpha,\nabla_{\beta\gamma}] &= C_{\alpha(\beta} W_{\gamma)},
\\
\label{gl:cc}
 [\nabla_{\alpha\beta},\nabla_{\gamma\delta}] &= -\frac{1}{2} C_{\alpha\gamma} F_{\beta\delta}
	-\frac{1}{2} C_{\alpha\delta} F_{\beta\gamma} 
	-\frac{1}{2} C_{\beta\delta} F_{\alpha\gamma}
	-\frac{1}{2} C_{\beta\gamma} F_{\alpha\delta},
\end{align}
\end{subequations}
where the field strengths are
\begin{subequations}
\begin{align}
\label{gl:ia}
 \Gamma_{\alpha\beta} &= -\frac{1}{2}\left( D_{(\alpha}\Gamma_{\beta)} -i \{\Gamma_\alpha,\Gamma_\beta\} \right),
\\
\label{gl:ib}
 W_\alpha &= -\frac{i}{2}D^\beta D_\alpha \Gamma_\beta - \frac{1}{2}[\Gamma^\beta,D_\beta\Gamma_\alpha]
  + \frac{i}{6}[\Gamma^\beta,\{\Gamma_\beta,\Gamma_\alpha\}],
&
 \nabla^\alpha W_{\alpha} &= 0,
\\
\label{gl:ic}
 F_{\alpha\beta} &= \frac{1}{2}\nabla_{(\alpha} W_{\beta)}.
\end{align}
\end{subequations}

We would also like to recall some well known identities, which we are going to use a lot in this paper. 
We have the super-Jacobi identity
\begin{equation}
   (-1)^{\tilde{a}\tilde{c}}[[a,b]_\pm,c]_\pm
 + (-1)^{\tilde{a}\tilde{b}}[[b,c]_\pm,a]_\pm
 + (-1)^{\tilde{b}\tilde{c}}[[c,a]_\pm,b]_\pm = 0, 
\end{equation}
where $[\cdot,\cdot]_\pm$ stands for the graded commutator and the tilde denotes the Grassmann parity.
Then we have the super-Leibniz rule (from which rules for integration by parts follow)
\begin{align}
 \mathcal{D}(ab) &= (\mathcal{D}a)b + (-1)^{\tilde{\mathcal{D}}\tilde{a}}a(\mathcal{D}b),
&
 \mathcal{D}[a,b]_\pm &= [\mathcal{D}a,b]_\pm + (-1)^{\tilde{\mathcal{D}}\tilde{a}}[a,\mathcal{D}b]_\pm.
\end{align}
where for $\mathcal{D}$ we may substitute the Grassmann odd derivatives $D_\alpha$, $\pp$ 
or the Grassmann even derivatives $\partial_{\alpha\beta}$. 
We are also going to extensively use the cyclic property of the trace
\begin{equation}
 \text{tr}(ab) = (-1)^{\tilde{a}\tilde{b}}\text{tr}(ba),
\end{equation}
and the identity
\begin{equation}
 \text{tr}(a[b,c]_\pm) = \text{tr}([a,b]_\pm c).
\end{equation}

\section{$SIM(1)$ Supersymmetry}

In this section, we are going to summarize some facts about the $SIM(1)$ supersymmetry. 
The detailed explanation of the $SIM(1)$ supersymmetry can be found in \cite{sim1}. 
Here, we are only going to mention some basic facts.

The $SIM(1)$ group is a subgroup of the Lorentz group consisting of all transformations that preserve 
a given null-direction, which means that a given null-vector is preserved up to a rescaling. 
This null-vector will be denoted $n$ and we will assume that it is chosen such that it has only one nonzero 
coordinate $n^{++}=1$.
When we work with spinors it is useful to write the Lorentz transformations with the help of the  group 
$SL(2,\mathbb{R})$, which is a double cover of $SO_+(2,1)$. The Lorentz transformation of a spinor is then
given as a multiplication by a matrix from $SL(2,\mathbb{R})$.
The reduction of the Lorentz group to the $SIM(1)$ subgroup corresponds to the reduction of the group 
$SL(2,\mathbb{R})$ to its two-dimensional subgroup of triangular matrices.
Thus, the $SIM(1)$ transformation of a general spinor $\psi$ is given as
\begin{align}\label{sptr}
 \begin{pmatrix} \psi'^+ \\ \psi'^- \end{pmatrix} &= 
 \begin{pmatrix} e^{-A}&-B \\ 0&e^{A} \end{pmatrix}
 \begin{pmatrix} \psi^+ \\ \psi^- \end{pmatrix}
&&\Leftrightarrow&
 \begin{pmatrix} \psi'_+ \\ \psi'_- \end{pmatrix} &= 
 \begin{pmatrix} e^{A}&0 \\ B&e^{-A} \end{pmatrix}
 \begin{pmatrix} \psi_+ \\ \psi_- \end{pmatrix},
\end{align}
where $A,B\in\mathbb{R}$.

The space of spinors $\mathcal{S}$ is not irreducible when the symmetry is reduced to the $SIM(1)$ subgroup.
There are two irreducible spaces that are important for understanding of the $SIM(1)$ supersymmetry.
The first one is the space $\mathcal{S}_{\text{invariant}}$ of all spinors that satisfy the condition $\fsl{n}\psi = 0$,
the second one is the quotient space $\mathcal{S}_{\text{quotient}} = \mathcal{S}/\mathcal{S}_{\text{invariant}}$.
The space $\mathcal{S}_{\text{invariant}}$ consists of spinors that have $\psi_+$ coordinate equal to zero,
the space $\mathcal{S}_{\text{quotient}}$ can be conveniently described if we choose in each equivalence class
a representative for which the coordinate $\psi_-$ vanish.
The $SIM(1)$ transformations change spinors from these spaces as
\begin{align}\label{sptri}
 \begin{pmatrix} 0 \\ \psi'_- \end{pmatrix} &= e^{-A} \begin{pmatrix} 0 \\ \psi_- \end{pmatrix},
&
 \left[ \begin{pmatrix} \psi'_+ \\ 0 \end{pmatrix} \right ] 
 &= e^{A} \left [ \begin{pmatrix} \psi_+ \\ 0 \end{pmatrix} \right].
\end{align}

The $SIM(1)$ supersymmetry is a reduction of the super-Poincare supersymmetry that we get if the Lorentz symmetry 
is reduced to its $SIM(1)$ subgroup. However, it is not enough to reduce the spacetime symmetry, the amount 
of supersymmetry has to be reduced as well. We keep only half of the $\mathcal{N}=1$ supersymmetry when 
we make the reduction to the $SIM(1)$ supersymmetry. Thus, we can say that the $SIM(1)$ supersymmetry describes 
$\mathcal{N}=1/2$ supersymmetry.
Since the amount of supersymmetry is halved, the number of anticommuting coordinates that parametrize  
$SIM(1)$ superspace has to be halved as well. 
So, in the $SIM(1)$ supersymmetry we have only one supercharge $S_+$, one anticommuting coordinate $\theta_-$ to which
corresponds the spinor derivative $\pp$. 
The supersymmetry generator and the spinor derivative can be written as
\begin{align}
 S_+ &= \partial_+ + i\theta_- \partial_{++},
&
 d_+ &= \partial_+ - i\theta_- \partial_{++}.
\end{align}
and they satisfy
\begin{align}
 \{S_+,S_+\} &= 2\partial_{++},
&
 \{S_+,d_+\} &= 0,
&
 \{d_+,d_+\} &= -2\partial_{++},
&
 \partial_+ \theta_- &= -i.
\end{align}
The anticommuting coordinate $\theta_-$ transforms under the $SIM(1)$ group as a spinor from $\mathcal{S}_{\text{invariant}}$, 
the supersymmetry generator and the spinor derivative transform under the $SIM(1)$ group as spinors from 
$\mathcal{S}_{\text{quotient}}$.

\section{ Chern-Simons Theory with $SIM(1)$ Projections} \label{cs1}

In most application of the Chern-Simons theory, like the BLG theory  \cite{1}-\cite{5} 
and the ABJM theory \cite{apjm}-\cite{abjm1}, the Chern-Simons theory is coupled to matter fields. 
We can write  the total action for a simple matter-Chern-Simons  theory as 
\begin{equation}
 S  = S_M + S_{CS},
\end{equation}
 where $S_M$  is the action for matter fields, and $S_{CS}$ is the action for the Chern-Simons theory. The action for the Chern-Simons theory in $\mathcal{N} = 1 $ superspace can be written as 
\begin{equation}\label{laction}
 S_{CS} = \frac{k}{4\pi} \text{tr} \int\di^3x D^2 \left( \Gamma^\alpha W_\alpha 
 - \frac{1}{6} \{ \Gamma^\alpha , \Gamma^\beta \} \Gamma_{\alpha\beta} \right),  
\end{equation}
where $k$ is  the level of the Chern-Simons theory.
The matter fields coupled to gauge fields  have already been studied in 
$SIM(1)$ superspace \cite{sim1}, and the analysis here will not be very different.
It is  possible  to break the Lorentz symmetry down to the $SIM(1)$ symmetry by adding a suitable mass term \cite{sim1}
\begin{equation}
 S_m = - m^2 \int \di^3x \nabla_+ \left( \phi^\dagger \frac{\nabla_+}{\nabla_{++}}\phi \right).
\end{equation}
 It may be noted that the mass deformed BLG theory is thought to be related to the theory of M5-branes \cite{blgmass}.
 In this paper, we will not analyse the detail construction of the BLG theory and the ABJM theory, 
but we will just assume that the Chern-Simons theory is suitably coupled to a term that breaks the Lorentz invariance to the $SIM(1)$ invariance. 
We will thus construct a Chern-Simons theory in $SIM(1)$ superspace with $\mathcal{N} = 1/2$ supersymmetry, 
and explicitly demonstrate that this theory is gauge invariant. 

So,   we are going to write the Chern-Simons action 
in $SIM(1)$ superspace. We are going to do this in two steps. In the first step, we are going to write 
the Chern-Simons action using $SIM(1)$ projections of $\Gamma_\alpha$, $\Gamma_{\alpha\beta}$ 
and $SIM(1)$ projections of field strengths $W_\alpha$, $F_{\alpha\beta}$.
In this case $SIM(1)$ projection means that we
set the anticommuting coordinate $\theta_+$, which was removed when we made the reduction to 
$SIM(1)$ superspace, to zero. In the second step, we are going to replace these projections with 
$SIM(1)$ superfields which will have nicer transformation properties with respect to the $SIM(1)$ group.
The gauge transformations will be very complicated for these $SIM(1)$ superfields.

Let us start with step one. We define $SIM(1)$ projections
\begin{align}\label{sdefg}
 \gp &= \Gamma_+ \vert_{\theta_+=0},
&
 \gm &= \Gamma_- \vert_{\theta_+=0},
\nonumber\\
 \gpp &= \Gamma_{++} \vert_{\theta_+=0},
&
 \gmp &= \Gamma_{+-} \vert_{\theta_+=0},
&
 \gmm &= \Gamma_{--} \vert_{\theta_+=0}.
\end{align}
of connections and $SIM(1)$ projections
\begin{align}\label{sdefs}
 \wp &= W_+ \vert_{\theta_+=0},
&
 \wm &= W_- \vert_{\theta_+=0},
\nonumber\\
 \fpp &= F_{++} \vert_{\theta_+=0},
&
 \fmp &= F_{+-} \vert_{\theta_+=0},
&
 \fmm &= F_{--} \vert_{\theta_+=0},
\end{align}
of field strengths.

In the $SIM(1)$ formulation of the gauge theory we will use the projections $\gp$, $\gm$, $\gpp$, $\gmp$ and $\gmm$.
We will see that this set of projections contains all the information we need. 
The only constraint that this set of projections has to satisfy is is
\begin{equation}\label{ssc}
 \pp\gp = -\gpp + \frac{i}{2}\{\gp,\gp\},
\end{equation}
which follows from $\{\nabla_+,\nabla_+\}=-2\nabla_{++}$.
The projections of the field strengths \eqref{sdefs} can be expressed as (anti)commutators of 
$\nabla_+\vert_{\theta_+=0}$, $\nabla_{++}\vert_{\theta_+=0}$, $\nabla_{+-}\vert_{\theta_+=0}$ 
and $\nabla_{--}\vert_{\theta_+=0}$
\begin{align}\label{wf_s}
 \wp &= i [ \nabla_+ , \nabla_{+-} ] \vert_{\theta_+=0} 
 = \pp \gmp - \pmp \gp -i [ \gp , \gmp ],
\nonumber\\
 \wm &= \frac{i}{2} [ \nabla_+ , \nabla_{--} ] \vert_{\theta_+=0} 
 = \frac{1}{2} \left( \pp \gmm - \pmm \gp -i [ \gp , \gmm ] \right),
\nonumber\\
 \fpp &=  -i [ \nabla_{++} , \nabla_{+-} ] \vert_{\theta_+=0} 
 = - \ppp \gmp + \pmp \gpp +i [ \gpp , \gmp ],
\nonumber\\
 \fmp &= -\frac{i}{2} [ \nabla_{++} , \nabla_{--} ] \vert_{\theta_+=0} 
 = \frac{1}{2} \left( - \ppp \gmm + \pmm \gpp +i [ \gpp , \gmm ] \right),
\nonumber\\
 \fmm &= -i [ \nabla_{+-} , \nabla_{--} ] \vert_{\theta_+=0} 
 = - \pmp \gmm + \pmm \gmp +i [ \gmp , \gmm ].
\end{align}
The projections of $D_-$ derivatives of the connections 
$D_-\Gamma_\alpha\vert_{\theta_+=0}$ and $D_-\Gamma_{\alpha\beta}\vert_{\theta_+=0}$ can be calculated as
\begin{align}\label{spmproj}
 ( D_- \Gamma_+ ) \vert_{\theta_+=0} &= - 2 \gmp - \pp \gm + i \{ \gp , \gm \},
\nonumber\\
 ( D_- \Gamma_- ) \vert_{\theta_+=0} &= - \gmm + \tfrac{i}{2} \{ \gm , \gm \},
\nonumber\\
 ( D_- \Gamma_{++} ) \vert_{\theta_+=0} &= \ppp \gm - 2 \wp + i [ \gm , \gpp ],
\nonumber\\
 ( D_- \Gamma_{+-} ) \vert_{\theta_+=0} &= \pmp \gm - \wm + i [ \gm , \gmp ],
\nonumber\\
 ( D_- \Gamma_{--} ) \vert_{\theta_+=0} &= \pmm \gm + i [ \gm , \gmm ].
\end{align}
The fact that we know projections of all connections and projections of their $D_-$ derivatives allows us to reconstruct
the original connections. A Lorentz superfield $\Phi$ can be written with the help of its projection 
$\Phi\vert_{\theta_+=0}$ and projection of its $D_-$ derivative $(D_-\Phi)\vert_{\theta_+=0}$ as
\begin{equation}
 \Phi = (\Phi\vert_{\theta_+=0}) 
 - i \theta_+ \left[ ((D_-\Phi)\vert_{\theta_+=0}) + i \theta_- \partial_{+-} (\Phi\vert_{\theta_+=0}) \right].
\end{equation}
The same thing can also be done in the case of connections.
The projection $\gm$ does not appear in any of our results and we do not need it to calculate
any of field strengths.

The infinitesimal gauge transformations of the projections \eqref{sdefg} are quite simple
\begin{align}\label{sgcon}
 \delta_g \gp &= i[k,\gp] + \pp k,
& 
 \delta_g \gm &= i[k,\gm] + \kappa_-,
\nonumber\\
 \delta_g \gpp &= i[k,\gpp] + \ppp k,
&
 \delta_g \gmp &= i[k,\gmp] + \pmp k,
&
 \delta_g \gmm &= i[k,\gmm] + \pmm k,
\end{align}
where $k$ is the $SIM(1)$ projection $k=K\vert_{\theta_+=0}$ of a real scalar superfield $K$
and $\kappa_-=(D_-K)\vert_{\theta_+=0}$ is a projection of its derivative.

The $SIM(1)$ projections of the field strengths \eqref{sdefs} transform covariantly,
thus their infinitesimal gauge transformations are just commutators with $k$
\begin{align}\label{sgfld}
 \delta_g \wp &= i[k,\wp],
&
 \delta_g \wm &= i[k,\wm],
\nonumber\\
 \delta_g \fpp &= i[k,\fpp],
&
 \delta_g \fmp &= i[k,\fmp],
&
 \delta_g \fmm &= i[k,\fmm].
\end{align}
The infinitesimal $SIM(1)$ transformations of superfields can be split into two parts, $\delta_s=\hat{\delta}_s+\delta'_s$.
The first part $\hat{\delta}_s$ corresponds to the change that can be attributed to the representation carried by 
the superfield, the second part $\delta'_s$ corresponds to the change caused by the transformation of the superspace 
coordinates. We will focus on the first part because the invariance of the action with respect to the second part
is ensured by integration over superspace. The infinitesimal $SIM(1)$ transformations of the projections
\eqref{sdefg}, \eqref{sdefs}
follow directly from \eqref{sptr}
\begin{align}\label{sst}
 \hat{\delta}_s \gp &= A\gp,
&
 \hat{\delta}_s \gm &= -A\gm + B\gp,
\nonumber\\
 \hat{\delta}_s \gpp &= 2A\gpp,
&
 \hat{\delta}_s \gmp &= B\gpp,
&
 \hat{\delta}_s \gmm &= -2A\gmm + 2B\gmp,
\nonumber\\
 \hat{\delta}_s \wp &= A\wp,
&
 \hat{\delta}_s \wm &= -A\wm + B\wp,
\nonumber\\
 \hat{\delta}_s \fpp &= 2A\fpp,
&
 \hat{\delta}_s \fmp &= B\fpp,
&
 \hat{\delta}_s \fmm &= -2A\fmm + 2B\fmp,
\end{align}
and for derivatives we have
\begin{align}\label{sstp}
 \delta_s\pp &= A\pp,
&
 \delta_s\ppp &= 2A\ppp,
&
 \delta_s\pmp &= B\ppp,
&
 \delta_s\pmm &= -2A\pmm + 2B\pmp.
\end{align}
 
\subsection{Chern-Simons Action with $SIM(1)$ Projections}

In this section, we are going to write down the Chern-Simons action \eqref{laction} in $SIM(1)$ superspace.
This $SIM(1)$ action should depend only on $SIM(1)$ superfields 
$\gamma_\alpha$, $\gamma_{\alpha\beta}$, $w_\alpha$, $f_{\alpha\beta}$ 
and their $\pp$, $\partial_{\alpha\beta}$ derivatives. 
Moreover, it  should be written as a $SIM(1)$ superspace integral of a $SIM(1)$ Lagrangian.
Since we do not have the $\theta^-$ coordinate in $SIM(1)$ superspace, there is no integration
over $\theta^-$ in the $SIM(1)$ superspace integral.

The first step to obtain the $SIM(1)$ action is the reduction of the integration measure. 
We expand the summation in the action and write everything with lower indices
\begin{multline}
 S_{CS} = \frac{k}{4\pi} \text{tr} \int\di^3x D_+ D_- \Big( 
 - \Gamma_+ W_- + \Gamma_- W_+
 \\
 -\frac{i}{6} \{\Gamma_+,\Gamma_+\}\Gamma_{--} 
 + \frac{i}{3} \{\Gamma_+,\Gamma_-\}\Gamma_{+-}
 -\frac{i}{6} \{\Gamma_-,\Gamma_-\}\Gamma_{++} 
 \Big),
\end{multline}
and then we integrate over $D_-$
\begin{multline}\label{sc1}
 S_{CS} = \frac{k}{4\pi} \text{tr} \int\di^3x D_+ \Big( 
 - (D_-\Gamma_+) W_- + \Gamma_+ (D_- W_-) + (D_-\Gamma_-) W_+ - \Gamma_- (D_- W_+)
 \\
 - \frac{i}{6} (D_- \{\Gamma_+,\Gamma_+\}) \Gamma_{--} -\frac{i}{6} \{\Gamma_+,\Gamma_+\}(D_- \Gamma_{--})
 + \frac{i}{3} (D_- \{\Gamma_+,\Gamma_-\}) \Gamma_{+-} 
 \\
 + \frac{i}{3} \{\Gamma_+,\Gamma_-\} (D_- \Gamma_{+-})
 - \frac{i}{6} (D_- \{\Gamma_-,\Gamma_-\}) \Gamma_{++} - \frac{i}{6} \{\Gamma_-,\Gamma_-\} (D_- \Gamma_{++}) 
 \Big).
\end{multline}
The integration measure can now be written as $\di^3x\pp$ and the expression inside the integral can be 
expressed with $SIM(1)$ superfields. 
If there is no $D_-$ derivative acting on the Lorentzian superfield, then we can replace it with its projection
\eqref{sdefg}, \eqref{sdefs}. If the $D_-$ derivative acts on one of the connections 
$\Gamma_\alpha$, $\Gamma_{\alpha\beta}$, then we use the substitutions \eqref{spmproj}.
In the case where the $D_-$ derivative acts on the field strength $W_\alpha$, we calculate the projection 
from \eqref{gl:ic}
\begin{align}
 D_- W_+ \vert_{\theta_+=0} &= \fmp + i \{ \gm , \wp \},
&
 D_- W_- \vert_{\theta_+=0} &= \fmm + i \{ \gm , \wm \}.
\end{align}
What remains is to calculate the projection of the $D_-$ derivative of the anticommutator 
$\{\Gamma_\alpha,\Gamma_\beta\}$. 
In the case of $\{\Gamma_+,\Gamma_+\}$ we get
\begin{multline}
 D_- \{ \Gamma_+ , \Gamma_+ \} \vert_{\theta_+=0}
 = -2 [ \Gamma_+ , D_- \Gamma_+ ] \vert_{\theta_+=0}
 = - 2 [ \gp , - 2 \gmp - \pp \gm + i \{ \gp , \gm \}]
 \\
 = 4 [ \gp ,\gmp ] - 2 \pp \{ \gp ,\gm \} - 2 [ \gpp , \gm ] + i [ \{ \gp, \gp\} , \gm ] + 2i [ \{ \gp , \gm \}, \gp ]
 \\
 = 4 [ \gp , \gmp ] - 2 \pp \{ \gp ,\gm \} - 2 [ \gpp , \gm ],
\end{multline}
where we used 
\begin{equation}
 [ \gp , \pp \gm ] 
 = - \pp \{ \gp , \gm \} + [ \pp \gp , \gm ] 
 = - \pp \{ \gp , \gm \} + [ - \gpp + \tfrac{i}{2} \{ \gp , \gp \} , \gm ],
\end{equation}
and the super-Jacobi identity $2[\{\gp,\gm\},\gp]+[\{\gp,\gp\},\gm]=0$.
In the case of $\{\Gamma_+,\Gamma_-\}$ we get
\begin{multline}
 D_- \{ \Gamma_+ , \Gamma_- \} \vert_{\theta_+=0}
 = \left( [ D_- \Gamma_+ , \Gamma_- ] - [ \Gamma_+ , D_- \Gamma_- ] \right) \vert_{\theta_+=0}
 \\
 = [ - 2 \gmp - \pp \gm + i \{ \gp , \gm \} , \gm ] - [ \gp , - \gmm + \tfrac{i}{2} \{ \gm , \gm \}]
 \\
 = 2 [ \gm , \gmp ] + [ \gp , \gmm ] - \tfrac{1}{2} \pp \{ \gm , \gm \},
\end{multline}
where we used the super-Jacobi identity $2[\{\gp,\gm\},\gp]+[\{\gp,\gp\},\gm]=0$ and the identity
$[\pp\gm,\gm]=\tfrac{1}{2}\pp\{\gm,\gm\}$.
Finally, in the case of $\{\Gamma_-,\Gamma_-\}$ we get
\begin{multline}
 D_- \{ \Gamma_- , \Gamma_- \}  \vert_{\theta_+=0}
 = - 2 [ \Gamma_- , D_- \Gamma_- ] \vert_{\theta_+=0}
 \\
 = - 2 [ \gm , - \gmm + \tfrac{i}{2} \{ \gm , \gm \} ]
 = 2 [ \gm , \gmm ].
\end{multline}
When we substitute all the above expressions in \eqref{sc1}, we obtain
\begin{multline}\label{sc2}
 S_{CS} = \frac{k}{4\pi} \text{tr} \int\di^3x \pp \Big( 
 2 \gmp \wm 
 + (\pp\gm) \wm 
 - i \{\gp,\gm\} \wm 
 + \gp \fmm 
 \\
 + i\gp \{\gm,\wm\}
 - \gmm \wp 
 + \tfrac{i}{2} \{\gm,\gm\} \wp 
 - \gm \fmp 
 -i \gm \{\gm,\wp\}
 \\
 - \tfrac{2i}{3} [\gp,\gmp] \gmm 
 - \tfrac{i}{3} [\gm,\gpp] \gmm 
 + \tfrac{i}{3} (\pp\{\gp,\gm\})\gmm
 \\
 - \tfrac{i}{6} \{\gp,\gp\} (\pmm\gm) 
 + \tfrac{1}{6}\{\gp,\gp\} [\gm,\gmm]
 + \tfrac{2i}{3} [\gm,\gmp] \gmp 
 \\
 + \tfrac{i}{3} [\gp,\gmm] \gmp 
 - \tfrac{i}{6} (\pp\{\gm,\gm\}) \gmp
 + \tfrac{i}{3} \{\gp,\gm\} (\pmp\gm) 
 \\
 - \tfrac{i}{3} \{\gp,\gm\} \wm 
 - \tfrac{1}{3} \{\gp,\gm\} [\gm,\gmp]
 - \tfrac{i}{3}  [\gm,\gmm] \gpp
 \\
 - \tfrac{i}{6} \{\gm,\gm\} (\ppp\gm) 
 + \tfrac{i}{3} \{\gm,\gm\} \wp
 + \tfrac{1}{6} \{\gm,\gm\} [\gm,\gpp]
 \Big).
\end{multline}
This result satisfies all the requirements that we expect from the $SIM(1)$ action. It is expressed as a $SIM(1)$
superspace integral and it is expressed with the help of $SIM(1)$ superfields and their derivatives.
This action can be further simplified if we use the fact that all surface terms vanish.
We are going to need the surface terms that are total $\pp$ derivatives
\begin{align}
 &\text{tr} \int\di^3x \pp \big( \pp (\gm \wm) \big)
 = \text{tr} \int\di^3x \pp \big( (\pp\gm)\wm - \gm \fmp - i\gm\{\gp,\wm\} \big),
\nonumber\\
 &\text{tr} \int\di^3x \pp \big( \pp(\{\gp,\gm\}\gmm) \big) 
 = \text{tr} \int\di^3x \pp \big( (\pp\{\gp,\gm\})\gmm 
 \nonumber\\ &\qquad\qquad\qquad
 + \{\gp,\gm\}(\pmm\gp) + 2\{\gp,\gm\}\wm +i \{\gp,\gm\}[\gp,\gmm] \big),
\nonumber\\
 &\text{tr} \int\di^3x \pp \big( \pp(\{\gm,\gm\}\gmp) \big) 
 = \text{tr} \int\di^3x \pp \big( (\pp\{\gm,\gm\})\gmp 
 \nonumber\\ &\qquad\qquad\qquad
 + \{\gm,\gm\}(\pmp\gp) + \{\gm,\gm\}\wp +i \{\gm,\gm\}[\gp,\gmp] \big),
\end{align}
and also some surface terms that are total spacetime derivatives
\begin{align}
 \text{tr} \int\di^3x \pp \big[ \pmm(\{\gp,\gp\}\gm) \big]
 &= \text{tr} \int\di^3x \pp \big[ 2\{\gp,\gm\}(\pmm\gp) + \{\gp,\gp\}(\pmm\gm) \big],
 \nonumber\\
 \text{tr} \int\di^3x \pp \big[ \pmp(\{\gm,\gm\}\gp) \big]
 &= \text{tr} \int\di^3x \pp \big[ 2\{\gp,\gm\}(\pmp\gm) + \{\gm,\gm\}(\pmp\gp) \big],
 \nonumber\\
 \text{tr} \int\di^3x \pp \big[ \tfrac{1}{3}\ppp(\{\gm,\gm\}\gm) \big]
 &= \text{tr} \int\di^3x \pp \big[ \{\gm,\gm\}(\ppp\gm) \big].
\end{align}
If appropriate multiples of these surface terms are added to the action \eqref{sc2},
all terms that contain derivatives are eliminated. 
This gives us the action
\begin{multline}
 S_{CS} = \frac{k}{4\pi} \text{tr} \int\di^3x \pp \Big( 
 2 \gmp \wm 
 + \gp \fmm
 - \gmm \wp 
 - i\gp[\gmp,\gmm]
 \\
 + \tfrac{1}{3} \{\gp,\gm\} [\gp,\gmm]
 + \tfrac{1}{6} \{\gp,\gp\} [\gm,\gmm]
 \\
 - \tfrac{1}{6} \{\gm,\gm\} [\gp,\gmp]
 - \tfrac{1}{3} \{\gp,\gm\} [\gm,\gmp]
 + \tfrac{1}{6} \{\gm,\gm\} [\gm,\gpp]
 \Big),
\end{multline}
which could also be written as
\begin{multline}
 S_{CS} = \frac{k}{4\pi} \text{tr} \int\di^3x \pp \Big( 
 2 \gmp \wm 
 + \gp \fmm
 - \gmm \wp 
 - i\gp[\gmp,\gmm]
 \\
 + \tfrac{1}{3} [\{\gp,\gm\},\gp]\gmm
 + \tfrac{1}{6} [\{\gp,\gp\},\gm]\gmm
 \\
 - \tfrac{1}{6} [\{\gm,\gm\},\gp]\gmp
 - \tfrac{1}{3} [\{\gp,\gm\},\gm]\gmp
 + \tfrac{1}{6} [\{\gm,\gm\},\gm]\gpp
 \Big).
\end{multline}
The terms that contain two (anti)commutators vanish because of the super-Jacobi identities
\begin{align}
2[\{\gp,\gm\},\gp]+[\{\gp,\gp\},\gm]&=0,
&
2[\{\gp,\gm\},\gm]+[\{\gm,\gm\},\gp]&=0
\end{align}
and the identity $[\{\gm,\gm\},\gm]=0$.
The final expression for the Chern-Simons action in $SIM(1)$ superspace is
\begin{equation}\label{saction}
 S_{CS} = \frac{k}{4\pi} \text{tr} \int\di^3x \pp \Big( 
 2 \gmp \wm 
 + \gp \fmm
 - \gmm \wp 
 - i\gp[\gmp,\gmm]
 \Big).
\end{equation}


\subsection{Gauge Invariance of the Action}

In this section, we are going to verify the gauge invariance of the $SIM(1)$ superspace Chern-Simons action 
\eqref{saction} that we have derived in the previous section. 
According to \eqref{sgcon}, \eqref{sgfld}, the change in the Chern-Simons action caused by an infinitesimal 
gauge transformation is
\begin{multline}
 \delta_g S_{CS} = \frac{k}{4\pi} \text{tr} \int\di^3x d_+ \Big(
   2 \left( i[k,\gmp] 
 + \pmp k \right) \wm 
 + 2 \gmp \left( i[k,\wm] \right)
 + \left( i[k,\gp] + \pp k \right) \fmm 
 \\
 + \gp \left( i[k,\fmm] \right)
 - \left( i[k,\gmm] + \pmm k \right) \wp 
 - \gmm \left( i[k,\wp] \right)
 -i \left( i[k,\gp] + \pp k \right) [\gmp,\gmm] 
 \\
 -i \gp \left[ i[k,\gmp] + \pmp k, \gmm \right] 
 -i \gp \left[ \gmp, i[k,\gmm] + \pmm k \right]
 \Big).
\end{multline}
As expected, all terms that contain commutator $i[k,\cdot]$ cancel due to the cyclic property of the trace and 
the super-Jacobi identity, so we are left with
\begin{multline}
 \delta_g S_{CS} = \frac{k}{4\pi} \text{tr} \int\di^3x d_+ \Big(
   2 (\pmp k) \wm 
 + (\pp k) \fmm 
 - (\pmm k) \wp
 \\
 -i (\pp k) [\gmp,\gmm] 
 -i \gp [\pmp k,\gmm] 
 -i \gp [\gmp,\pmm k]
 \Big).
\end{multline}
Now, we integrate by parts to move the derivatives away from the superfield $k$
\begin{equation}\begin{split}
 \delta_g S_{CS} = \frac{k}{4\pi} \text{tr} \int\di^3x d_+ \Big[ & k \Big(
 - 2 \pmp \wm 
 - \pp \fmm 
 + \pmm \wp
 \\
 &+i \pp [ \gmp , \gmm ] 
 -i \pmp [ \gp , \gmm ] 
 +i \pmm [ \gp , \gmp ]
 \Big) \Big]
 \\
 + \frac{k}{4\pi} \text{tr} \int\di^3x d_+ \Big[ &
   \pp \left( k \fmm -i k [ \gmp , \gmm ] \right)
 \\
 &+ \pmp \left( 2 k \wm +i k [ \gp , \gmm ] \right)
 + \pmm \left( - k \wp -i k [ \gp , \gmp ] \right)
 \Big].
\end{split}\end{equation}
The above expression consists of two parts, the bulk part and the surface part.
When we substitute for the field strengths \eqref{wf_s} and their derivatives
\begin{align}
 \pmm \wp &= \pmm \pp \gmp - \pmm \pmp \gp -i \pmm [ \gp , \gmp ] ,
\nonumber\\
 \pmp \wm &= \frac{1}{2}\left( \pmp \pp \gmm - \pmp \pmm \gp -i \pmp [ \gp , \gmm ] \right) ,
\nonumber\\
 \pp \fmm &= - \pmp \pp \gmm + \pmm \pp \gmp +i \pp [ \gmp , \gmm ],
\end{align}
then the terms in the bulk part cancel one another and only the surface term remains
\begin{multline}
 \delta_g S_{CS} = \frac{k}{4\pi} \text{tr} \int\di^3x d_+ \Big[
   \pp \left( - k ( \pmp \gmm ) + k ( \pmm \gmp) \right)
 \\
 + \pmp \left( k ( \pp \gmm ) - k ( \pmm \gp ) \right)
 + \pmm \left( - k ( \pp \gmp ) + k ( \pmp \gp ) \right)
 \Big].
\end{multline}
This surface term vanishes because we assume that we are working on a manifold without a boundary, so our action is gauge invariant.

\subsection{$SIM(1)$ Invariance of the Action}

In the Lorentz invariant setting, the Lorentz invariance is manifest. It follows from the fact that upper indices are 
summed with lower indices. In the $SIM(1)$ setting, the invariance under $SIM(1)$ transformations is not manifest,
its verification is a nontrivial task. 
In this section, we are going to verify the $SIM(1)$ invariance of the Chern-Simons action \eqref{saction}.

According to \eqref{sst} and \eqref{sstp}, the infinitesimal change of the action is
\begin{multline}
 \delta_s S_{CS} = \frac{k}{4\pi} \text{tr} \int\di^3x \pp \Big(
 A( 2 \gmp\wm + \gp\fmm - \gmm\wp -i \gp[\gmp,\gmm] )
 \\
 + 2B \gpp\wm -2A \gmp\wm +2B \gmp\wp
 +A \gp\fmm -2A \gp\fmm +2B \gp\fmp
 \\
 +2A \gmm\wp -2B \gmp\wp -A \gmm\wp
 -iA \gp[\gmp,\gmm] 
 \\
 -iB \gp[\gpp,\gmm] +2iA \gp[\gmp,\gmm] -2iB \gp[\gmp,\gmp]
 \Big).
\end{multline}
The terms with $A$ cancel one another because each term (including $\pp$ from the integration measure) 
contains the same number of plus and minus indices, so, for each $+A$ term there is one $-A$ term.
Thus, we are left with
\begin{equation}
 \delta_s S_{CS} = B \frac{k}{4\pi} \text{tr} \int\di^3x \pp \Big(
 2 \gpp\wm +2 \gp\fmp -i \gp[\gpp,\gmm]
 \Big).
\end{equation}
After we substitute for the field strengths \eqref{wf_s}, we obtain
\begin{multline}\label{ssi1}
 \delta_s S_{CS} = B \frac{k}{4\pi} \text{tr} \int\di^3x \pp \Big(
   \gpp ( \pp \gmm )
 - \gpp ( \pmm \gp )
 - \gp ( \ppp \gmm )
 \\
 + \gp ( \pmm \gpp )
 + i [ \gp, \gpp ] \gmm
 \Big).
\end{multline}
We will show that the above expression can be written as a surface term. In order to do that we will need the
following total $\pp$ derivatives
\begin{multline}
 \text{tr} \big[ \pp \big( \gp ( \pp \gmm ) \big) \big] 
 = \text{tr} \big[ 
 ( \pp \gp ) ( \pp \gmm ) - \gp ( \pp \pp \gmm )
 \big]
 \\
 = \text{tr} \big[ 
 - \gpp ( \pp \gmm ) 
 + \tfrac{i}{2} \{ \gp , \gp \} ( \pp \gmm )
 + \gp ( \ppp \gmm )
 \big] ,
\end{multline}
and 
\begin{multline}
 \text{tr} \big[ \pp \big( \gp ( \pmm \gp ) \big) \big] 
 = \text{tr} \big[ 
 ( \pp \gp ) ( \pmm \gp ) - \gp ( \pmm \pp \gp )
 \big]
 \\
 = \text{tr} \big[ 
 - \gpp ( \pmm \gp ) 
 + \tfrac{i}{2} \{ \gp , \gp \} ( \pmm \gp )
 + \gp ( \pmm \gpp )
 - \tfrac{i}{2} \gp ( \pmm \{ \gp , \gp \} )
 \big]
 \\
 = \text{tr} \big[ 
 - \gpp ( \pmm \gp ) 
 + \gp ( \pmm \gpp )
 - \tfrac{i}{6} \pmm ( \gp \{ \gp , \gp \} )
 \big].
\end{multline}
Here we used \eqref{ssc}, the identity $\pp\pp=-\ppp$ and
\begin{equation}
 \text{tr} \big[ \pmm ( \gp \{ \gp , \gp \} ) \big] 
 = 3 \text{tr} \big[ (\pmm \gp ) \{ \gp , \gp \} \big]
 = \frac{3}{2} \text{tr} \big[ \gp ( \pmm \{ \gp , \gp \} ) \big] .
\end{equation}
Now, the infinitesimal change of the action \eqref{ssi1} can be written as
\begin{multline}
 \delta_s S_{CS} = B \frac{k}{4\pi} \text{tr} \int\di^3x \pp \Big(
 - \pp \big( \gp ( \pp \gmm ) \big)
 + \pp \big( \gp ( \pmm \gp ) \big)
 \\
 + \frac{i}{2} \{ \gp , \gp \} ( \pp \gmm )
 + \frac{i}{6} \pmm ( \gp \{ \gp , \gp \} )
 + \frac{i}{2} ( \pp \{ \gp , \gp \} ) \gmm
 \Big),
\end{multline}
where we also used the identity
\begin{equation}\label{ppgpgp}
 \pp \{ \gp , \gp \} = -2 [ \gp, \pp \gp ]
 = 2 [\gp , \gpp ] -i [ \gp , \{ \gp, \gp \} ] = 2  [\gp , \gpp ].
\end{equation}
The above expression for $\delta_s S_{CS}$ can be written as a sum of a total $\pmm$ derivative and a total 
$\pp$ derivative 
\begin{multline}
 \delta_s S_{CS} = B \frac{k}{4\pi} \text{tr} \int\di^3x \pp \Big[
 \pmm \Big( \frac{i}{6} \gp \{ \gp , \gp \} \Big)
 \\
 + \pp \Big( - \gp ( \pp \gmm ) +  \gp ( \pmm \gp ) + \frac{i}{2} \{ \gp , \gp \} \gmm \Big)
 \Big].
\end{multline}
This term vanishes since we assume that we are working on a manifold without a boundary, so the Chern-Simons action \eqref{saction} is $SIM(1)$ invariant.

\section{Chern-Simons Theory with Redefined  Superfields}

The superfields that have been used to write down the action \eqref{saction} have nontrivial $SIM(1)$ transformation 
properties. The matrix that appear in \eqref{sptr} is not diagonal, so the $SIM(1)$ transformations 
mix the superfields with minus indices with those that have plus indices.
However, it is possible to redefine  $SIM(1)$ superfields such that they they do not suffer
from this deficiency \cite{sim1}.
The $SIM(1)$ transformation properties of these redefined  superfields are simpler because they do not carry 
representation of $\mathcal{S}$, they carry representation of $S_{\text{invariant}}$ and 
$S_{\text{quotient}}$ instead.

The idea behind these redefined $SIM(1)$ superfields could be most easily understood on an example of a spinor superfield. 
Let us consider a spinor superfield $\Psi$ and its projections
\begin{align}
 \psi_+ &= \Psi_+ \vert_{\theta_+=0},
&
 \psi_- &= \Psi_- \vert_{\theta_+=0},
\end{align}
which transform under the infinitesimal $SIM(1)$ transformations as
\begin{align}\label{simtmp}
 \hat{\delta}_s\psi_+ &= A \psi_+,
&
 \hat{\delta}_s\psi_- &= - A \psi_+ + B\psi_+.
\end{align}
The transformation rule for the projection $\psi_+$ is  a consequence of the fact that 
the projection $\psi_+$ carries the same representation that we have on the space $S_{\text{invariant}}$ 
\eqref{sptri}.
The $SIM(1)$ transformation of the $\psi_-$ projection mixes it with $\psi_+$ projection. 
Instead of the $SIM(1)$ superfield $\psi_-$ we are going to use the redefined $SIM(1)$ superfield 
 $\psi_\times$ that carries the same representation as we have on $S_{\text{quotient}}$ \eqref{sptri}.
For this purpose we define the operator
\begin{equation}
 \partial_{\times\alpha} = \frac{\partial_{+\alpha}}{\ppp},
\end{equation}
or in components
\begin{align}\label{pxmdef}
 \partial_{\times+} &= 1,
&
 \pxm &= \frac{\pmp}{\ppp}.
\end{align}
The superfield $\psi_\times$ is defined with the help of this operator as
\begin{equation}
 \psi_\times = i\partial_\times{}^\alpha\Psi_\alpha \vert_{\theta_+=0} = \psi_- - \pxm \psi_+.
\end{equation}
The infinitesimal $SIM(1)$ transformation of this superfield, does not mix it with $\psi_+$,
\begin{equation}\label{simtx}
 \hat{\delta}_s\psi_\times = -A \psi_\times.
\end{equation}
The $SIM(1)$ superfields that we are going to use in this section will be defined in exactly the same way. 
Although the above example used a superfield that carries only one index, the procedure that has been described
can be easily generalized for superfields that carry more than one index. If there are multiple indices, we just
need to repeat the above procedure for each index.

Instead of $SIM(1)$ superfields $\gp$, $\gm$, $\gpp$, $\gmp$, $\gmm$ that has been used to describe the Chern-Simons
in section \ref{cs1}, we are going to use superfields $\gp$, $\gpp$ that were defined in \eqref{sdefg} and
\begin{align}\label{gxdef}
 \gx &= i \left( \pxu{\alpha} \Gamma_\alpha \right) \vert_{\theta_+=0}
 = \gm - \pxm \gp ,
\nonumber\\
 \gxp &= i \left( \pxu{\alpha} \Gamma_{\alpha+} \right) \vert_{\theta_+=0} 
 = \gmp - \pxm \gpp ,
\nonumber\\
 \gxx &= -\left( \pxu{\alpha}\pxu{\beta} \Gamma_{\alpha\beta} \right)\vert_{\theta_+=0} 
 = \gmm -2 \pxm \gmp + \pxm^2 \gpp.
\end{align}
The infinitesimal gauge transformations of these superfields are more complicated that the ones for 
$\gm$, $\gmp$, $\gmm$,
\begin{align}
 \delta_g \gx &= i[k,\gx] - \pxu{\alpha}[k,\pxd{\alpha}\gp] + \kx,
\nonumber\\
 \delta_g \gxp &= i[k,\gxp] - \pxu{\alpha}[k,\pxd{\alpha}\gpp],
\nonumber\\
 \delta_g \gxx &= i[k,\gxx] - 2\pxu{\alpha}[k,\pxd{\alpha}\gxp] -i \pxu{\alpha}\pxu{\beta}[k,\pxd{\alpha}\pxd{\beta}\gpp] + \pxx k,
\end{align}
where we use the redefined $SIM(1)$ superfield
\begin{equation}
\kx = i\left( \pxu{\alpha} D_\alpha K \right)\vert_{\theta_+=0} = \km - \pxm \pp k,
\end{equation}
instead of $\km$.
The $SIM(1)$ transformations of redefined superfields are very simple
\begin{align}
 \hat{\delta}_s \gp &= A\gp,
&
 \hat{\delta}_s \gpp &= 2A\gpp,
&
 \hat{\delta}_s \gxp &= 0,
&
 \hat{\delta}_s \gxx &= -2A\gxx.
\end{align}
In fact, we can treat $\times$ as a new type of index. Each of indices $+$, $-$ and $\times$ have a specific
transformation law \eqref{simtmp}, \eqref{simtx} attached to it. The action of the $SIM(1)$ group on $SIM(1)$ superfield
is  fully determined by the types of indices it carries.

\subsection{Chern-Simons Action with Redefined  Superfields }

In this section, we are going to rewrite the action \eqref{saction} such that it is expressed with the help of redefined
$SIM(1)$ superfields that have been introduced in the previous section.

Before we start our calculations, we would like to mention few properties of the operator $\pxm$ that we are going to 
need. There are two identities that follow directly from the definition of this operator
\begin{align}\label{ccip}
 \pmm &= \frac{\Box}{\ppp} + \ppp\pxm^2,
&
 \pmp &= \ppp\pxm,
\end{align}
and there is a rule for integration by parts 
\begin{equation}
 \int\di^3x \left( (\pxm f)g \right) = \int\di^3x \left( f(\pxm g) \right).
\end{equation}

In order to eliminate $\gmp$, $\gmm$ from the action \eqref{saction}, we have to make the substitutions
\begin{align}
 \gmp &= \gxp + \pxm\gpp,
&
 \gmm &= \gxx + 2\pxm\gxp + \pxm^2\gpp,
\end{align}
which are just inverse relations to \eqref{gxdef}.
We also have to substitute for the field strengths $\wp$ and $\wm$ \eqref{wf_s}
\begin{equation}
 \wp = \pp \gxp -i [ \gp , \gxp] + \pxu{\alpha} [ \gp , \pxd{\alpha} \gpp ],
\end{equation}
where 
\begin{equation}
 \pxu{\alpha}[\gp,\pxd{\alpha}\gpp] = i \pxm[\gp,\gpp] -i [\gp,\pxm\gpp],
\end{equation}
and
\begin{multline}
 \wm = \tfrac{1}{2} \pp\gxx - \tfrac{i}{2}[\gp,\gxx] 
 - \tfrac{1}{2} \pxx\gp
 + \pxm\pp\gxp
 \\
 -i [\gp,\pxm\gxp]
 + \tfrac{i}{2} \pxm^2[\gp,\gpp] - \tfrac{i}{2} [\gp,\pxm^2\gpp].
\end{multline}
The field strength $\fmm$ is given by
\begin{multline}
 \fmm = \pxx\gxp + i[\gxp,\gxx] - [\pxu{\alpha}\gxp,\pxd{\alpha}\gxp] 
 - \ppp\pxm\gxx 
 \\
 - \ppp\pxm^2\gxp + \pxx\pxm\gpp +i [\gxp,\pxm^2\gpp] 
 \\
 + i [\pxm\gpp,\gxx] + 2i [\pxm\gpp,\pxm\gxp] +i [\pxm\gpp,\pxm^2\gpp],
\end{multline}
where 
\begin{equation}
 [\pxu{\alpha}\gxp,\pxd{\alpha}\gxp]=-2i[\gxp,\pxm\gxp].
\end{equation}

Now, we are ready to evaluate each of the four terms that appear in the action \eqref{saction}.
The first term is
\begin{multline}
 2 \gmp \wm = 
 \exl{A1}{ \gxp(\pp\gxx) } 
 \exl{A2}{ -i \gxp[\gp,\gxx] }
 \exl{A3}{ - \gxp\left(\pxx\gp\right) }
 \exl{A4}{ +2 \gxp(\pxm\pp\gxp) }
 \\
 \exl{A5}{ -2i \gxp[\gp,\pxm\gxp] }
 \exl{A6}{ +i \gxp(\pxm^2[\gp,\gpp]) }
 \exl{A7}{ -i \gxp[\gp,\pxm^2\gpp] }
 \exl{A8}{ + (\pxm\gpp)(\pp\gxx) } 
 \\
 \exl{A9}{ -i (\pxm\gpp)[\gp,\gxx] }
 \exl{A10}{ - (\pxm\gpp)\left(\pxx\gp\right) }
 \exl{A11}{ +2 (\pxm\gpp)(\pxm\pp\gxp) }
 \\
 \exl{A12}{ -2i (\pxm\gpp)[\gp,\pxm\gxp] }
 \exl{A13}{ +i (\pxm\gpp)(\pxm^2[\gp,\gpp]) }
 \exl{A14}{ -i (\pxm\gpp)[\gp,\pxm^2\gpp] },
\end{multline}
the second term is
\begin{multline}
 2 \gp \fmm = 
 \exl{B1}{ \gp\left(\pxx\gxp\right) } 
 \exl{B2}{ +i \gp[\gxp,\gxx] }
 \exl{B3}{ - \gp[\pxu{\alpha}\gxp,\pxd{\alpha}\gxp] }
 \\
 \exl{B4}{ - \gp(\ppp\pxm\gxx) }
 \exl{B5}{ - \gp(\ppp\pxm^2\gxp) }
 \exl{B6}{ + \gp\left(\pxx\pxm\gpp\right) }
 \exl{B7}{ +i \gp[\gxp,\pxm^2\gpp] }
 \\
 \exl{B8}{ +i \gp[\pxm\gpp,\gxx] } 
 \exl{B9}{ +2i \gp[\pxm\gpp,\pxm\gxp] }
 \exl{B10}{ +i \gp[\pxm\gpp,\pxm^2\gpp] },
\end{multline}
the third term is
\begin{multline}
 - \gmm \wp = 
 \exl{C1}{ - \gxx(\pp\gxp) } 
 \exl{C2}{ +i \gxx[\gp,\gxp] }
 \exl{C3}{ - \gxx(\pxu{\alpha}[\gp,\pxd{\alpha}\gpp]) }
 \\
 \exl{C4}{ -2 (\pxm\gxp)(\pp\gxp) }
 \exl{C5}{ +2i (\pxm\gxp)[\gp,\gxp] }
 \exl{C6}{ +2i (\pxm\gxp)[\gp,\pxm\gpp] }
 \\
 \exl{C7}{ -2i (\pxm\gxp)(\pxm[\gp,\gpp]) }
 \exl{C8}{ - (\pxm^2\gpp)(\pp\gxp) }
 \exl{C9}{ +i (\pxm^2\gpp)[\gp,\gxp] }
 \\
 \exl{C10}{ +i (\pxm^2\gpp)[\gp,\pxm\gpp] }
 \exl{C11}{ -i (\pxm^2\gpp)(\pxm[\gp,\gpp]) },
\end{multline}
and finally the fourth term is
\begin{multline}
 -i \gp [ \gmp , \gmm ] = 
 \exl{D1}{ -i \gp[\gxp,\gxx] } 
 \exl{D2}{ -2i \gp[\gxp,\pxm\gxp] }
 \exl{D3}{ -i \gp[\gxp,\pxm^2\gpp] }
 \\
 \exl{D4}{ -i \gp[\pxm\gpp,\gxx] } 
 \exl{D5}{ -2i \gp[\pxm\gpp,\pxm\gxp] }
 \exl{D6}{ -i \gp[\pxm\gpp,\pxm^2\gpp] }.
\end{multline}
We placed a designation consisting of an uppercase letter followed by a number under each term on 
the right side so we can easily identify them in later calculations. 
We are going to use the equivalence sign to denote that expressions differ by something that has 
trace equal to a surface term
\begin{equation}
 f \sim g 
 \qquad \Leftrightarrow \qquad 
 \text{tr} \int\di^3x d_+ f = \text{tr}\int\di^3x d_+ g.
\end{equation}
The sum of the above terms constitute the integrand of the Chern-Simons action.
We are going to split this sum into smaller pieces, such that each piece is separately $SIM(1)$ invariant.
The terms 
\begin{equation}\label{cb1a}
 \exl{A3}{ - \gxp\left(\pxx\gp\right) } 
 \exl{B1}{ + \gp\left(\pxx\gxp\right) }
\end{equation}
will be kept as they are, the terms 
\begin{equation}\label{cb1b}
 \exl{A4}{ 2 \gxp(\pxm\pp\gxp) } 
 \exl{C4}{ -2 (\pxm\gxp)(\pp\gxp) } 
 \sim 
 0,
\end{equation}
and we also have
\begin{align}\label{cb1c}
 &\exl{A2}{ -i \gxp[\gp,\gxx] } 
 \exl{B2}{ +i \gp[\gxp,\gxx] } 
 \exl{C2}{ +i \gxx[\gp,\gxp] } 
 \exl{D1}{ -i \gp[\gxp,\gxx] } 
 \sim 
 2i \gp[\gxp,\gxx],
\nonumber\\
 &\exl{A1}{ \gxp(\pp\gxx) } 
 \exl{C1}{ - \gxx(\pp\gxp) } 
 \sim 
 2 \gxp(\pp\gxx).
\end{align}
Another expressions that are easy to handle are
\begin{multline}\label{cb2}
 \exl{A5}{ -2i \gxp[\gp,\pxm\gxp] } 
 \exl{B3}{ - \gp[\pxu{\alpha}\gxp,\pxd{\alpha}\gxp] } 
 \exl{C5}{ +2i (\pxm\gxp)[\gp,\gxp] }
 \\
 \exl{D2}{ -2i \gp[\gxp,\pxm\gxp] } 
 \sim 
 4i \gp[\gxp,\pxm\gxp] \sim -2\gp[\pxu{\alpha}\gxp,\pxd{\alpha}\gxp],
\end{multline}
and
\begin{multline}\label{cb3}
 \exl{A8}{ (\pxm\gpp)(\pp\gxx) }
 \exl{A9}{ -i (\pxm\gpp)[\gp,\gxx] }
 \exl{B4}{ - \gp(\ppp\pxm\gxx) }
 \\
 \exl{B8}{ +i \gp[\pxm\gpp,\gxx] }
 \exl{C3}{ - \gxx(\pxu{\alpha}[\gp,\pxd{\alpha}\gpp]) }
 \exl{D4}{ -i \gp[\pxm\gpp,\gxx] } 
 \\
 \sim
 2i \gp[\pxm\gpp,\gxx] -i \gp[\gpp,\pxm\gxx] + (-\pp\gpp+\ppp\gp)(\pxm\gxx)
 \\
 \sim 
 2i \gp[\pxm\gpp,\gxx] -2i \gp[\gpp,\pxm\gxx] \sim 2 \gp[\pxu{\alpha}\gpp,\pxd{\alpha}\gxx],
\end{multline}
where we used the identity \eqref{sgcon}, \eqref{ppgpgp}
\begin{equation}\label{sgip2}
 \pp \gpp = \ppp \gp + i [ \gp , \gpp ].
\end{equation}
We also have
\begin{multline}\label{cb4}
 \exl{A6}{ i \gxp(\pxm^2[\gp,\gpp]) }
 \exl{A7}{ -i \gxp[\gp,\pxm^2\gpp] }
 \exl{A11}{ +2 (\pxm\gpp)(\pxm\pp\gxp) }
 \\
 \exl{A12}{ -2i (\pxm\gpp)[\gp,\pxm\gxp] }
 \exl{B5}{ - \gp(\ppp\pxm^2\gxp) }
 \exl{B7}{ +i \gp[\gxp,\pxm^2\gpp] }
 \\
 \exl{B9}{ +2i \gp[\pxm\gpp,\pxm\gxp] }
 \exl{C6}{ +2i (\pxm\gxp)[\gp,\pxm\gpp] }
 \\
 \exl{C7}{ -2i (\pxm\gxp)(\pxm[\gp,\gpp]) }
 \exl{C8}{ - (\pxm^2\gpp)(\pp\gxp) }
 \\
 \exl{C9}{ +i (\pxm^2\gpp)[\gp,\gxp] }
 \exl{D3}{ -i \gp[\gxp,\pxm^2\gpp] }
 \exl{D5}{ -2i \gp[\pxm\gpp,\pxm\gxp] } 
 \\
 \sim
 -2i \gp[\pxm^2\gpp,\gxp] + 4i \gp[\pxm\gpp,\pxm\gxp] -2i \gp[\gpp,\pxm^2\gxp] 
 \\
 \sim
 2i\gp[\pxu{\alpha}\pxu{\beta}\gpp,\pxd{\alpha}\pxd{\beta}\gxp],
\end{multline}
where we used \eqref{sgip2} to show that
\begin{multline}
 \exl{A11}{ 2 (\pxm\gpp)(\pxm\pp\gxp) }
 \exl{B5}{ - \gp(\ppp\pxm^2\gxp) }
 \exl{C8}{ - (\pxm^2\gpp)(\pp\gxp) }
 \\
 \sim 
 (-\pp\gpp+\ppp\gp)(\pxm^2\gxp)
 \sim
 -i \gp[\gpp,\pxm^2\gxp].
\end{multline}
The terms that we have not used yet could be arranged into expressions
\begin{equation}\label{cc1}
 \exl{A10}{ - (\pxm\gpp)\left(\pxx\gp\right) }
 \exl{B6}{ + \gp\left(\pxx\pxm\gpp\right) } 
 \sim
 2 \gp\left(\pxx\pxm\gpp\right)
\end{equation}
and
\begin{multline}\label{cc6}
 \exl{A13}{ i (\pxm\gpp)(\pxm^2[\gp,\gpp]) }
 \exl{A14}{ -i (\pxm\gpp)[\gp,\pxm^2\gpp] }
 \exl{B10}{ +i \gp[\pxm\gpp,\pxm^2\gpp] }
 \\
 \exl{C10}{ +i (\pxm^2\gpp)[\gp,\pxm\gpp] }
 \exl{C11}{ -i (\pxm^2\gpp)(\pxm[\gp,\gpp]) }
 \exl{D6}{ -i \gp[\pxm\gpp,\pxm^2\gpp] }
 \\
 \sim
 2i \gp[\pxm\gpp,\pxm^2\gpp].
\end{multline}
These two expressions do not lead to contributions to the action that are separately $SIM(1)$ invariant, however, we can combine them into a $SIM(1)$ invariant expression.
In \eqref{cc1}, we use \eqref{ssc} to replace $\gpp$ with expression that contains only $\gp$, 
\begin{equation}\label{cc2}
 2 \gp\left(\pxx\pxm\gpp\right) = -2 \gp\left(\pxx\pxm\pp\gp\right) +i \gp\left(\pxx\pxm\{\gp,\gp\}\right).
\end{equation}
Next we will show that the first term on the right hand side is a surface term. We integrate by parts 
to move all derivatives and the propagator such that they act on the first term, and then we use the cyclic property 
of the trace to show that
\begin{equation}
 \gp\left(\pxx\pxm\gpp\right) 
 \sim - \left(\pxx\pxm\pp\gp\right)\gp 
 \sim - \gp\left(\pxx\pxm\pp\gp\right) 
 \sim 0.
\end{equation}
 In the second term on the right hand side of \eqref{cc2} we are going to use the identity \eqref{ccip} to express 
$\pxx$ with $\pmm$, $\pxm$ and $\ppp$,  
then we are going to use integration by parts to move some derivatives and propagators.  Finally, we are going to use the identity \eqref{ccip}
again, this time to get $\pxx$ back into our expression  
\begin{align}\label{cc3}
 &i \gp\left(\pxx\pxm\{\gp,\gp\}\right)
 \sim i(\pxm\gp)(\pmm\{\gp,\gp\} - \pxm^2\ppp\{\gp,\gp\})
 \nonumber\\
 &\sim 2i (\pxm\gp)\{\pmm\gp,\gp\} -2i (\pxm^3\gp)\{\ppp\gp,\gp\}
 \nonumber\\
 &\sim 2i (\pxm\gp)\left\{\pxx\gp,\gp\right\} +2i (\pxm\gp)\{\ppp\pxm^2\gp,\gp\} -2i (\pxm^3\gp)\{\ppp\gp,\gp\}.
\end{align}
Now we take two thirds of the second term on the right hand side of \eqref{cc2}
and two thirds of the expression that we obtained in \eqref{cc3}
\begin{align}\label{cc4}
 2 \gp\left(\pxx\pxm\gpp\right) 
 &\sim - \tfrac{2i}{3} \left(\pxx\gp\right)(\pxm\{\gp,\gp\}) + \tfrac{2i}{3} \left(\pxx\gp\right)\{\gp,\pxm\gp\} 
 \nonumber\\
 &\quad + \tfrac{2i}{3} (\pxm\gp)\{\ppp\pxm^2\gp,\gp\} -\tfrac{2i}{3} (\pxm^3\gp)\{\ppp\gp,\gp\}.
\end{align}
The first two terms can be written as
\begin{equation}\label{cb5}
 -\tfrac{2}{3} \left(\pxx\gp\right)(\pxu{\alpha}\{\gp,\pxd{\alpha}\gp\})
\end{equation}
which, when traced and integrated over $SIM(1)$ superspace, gives $SIM(1)$ invariant contribution to the action.
The third term in \eqref{cc4} is a surface term. In order to show that we use the identity
\begin{equation}\label{tada}
 0 \sim \ppp ((\pxm\gp)\{\pxm\gp,\pxm\gp\}) \sim 3(\ppp\pxm\gp)\{\pxm\gp,\pxm\gp\},
\end{equation}
and the identity
\begin{multline}\label{tada1}
 0 \sim \pmp (\gp\{\pxm\gp,\pxm\gp\})
\\
   \sim (\ppp\pxm\gp)\{\pxm\gp,\pxm\gp\} + 2 (\ppp\pxm^2\gp)\{\gp,\pxm\gp\}.
\end{multline}
To prove this identity, we used the second identity from \eqref{ccip}.
The first term in \eqref{tada1} vanishes according to identity \eqref{tada}, so the second term, which is the same as the third term in \eqref{cc4}, must vanish,
\begin{equation}
 (\ppp\pxm^2\gp)\{\gp,\pxm\gp\} \sim 0.
\end{equation}
As for the last term in \eqref{cc4}, we have the identity
\begin{multline}\label{cc5}
 0 \sim \pp\left( (\pp\{\gp,\gp\})(\pxm^3\gp) \right) 
 \\
 = -(\ppp\{\gp,\gp\})(\pxm^3\gp) - (\pp\{\gp,\gp\})(\pxm^3\pp\gp)
 \\
 = -2\{\ppp\gp,\gp\}(\pxm^3\gp) +2 [\gp,\gpp](\pxm^3\gpp) - \tfrac{i}{2}(\pp\{\gp,\gp\})(\pxm^3\{\gp,\gp\}),
\end{multline}
where we used \eqref{ssc} and \eqref{ppgpgp}. 
The last term $(\pp\{\gp,\gp\})(\pxm^3\{\gp,\gp\})$ in \eqref{cc5} is a surface term. 
In order to prove it, we use the cyclic property of the trace to exchange $\pp\{\gp,\gp\}$ with $\pxm^3\{\gp,\gp\}$, integrate by parts to move $\pp$ from the first anticommutator to the second anticommutator and to move $\pxm^3$ from the second anticommutator to the first anticommutator.
The result that we obtain in this way is the same as the original expression, but with opposite sign, hence it must be a surface term.
Thus \eqref{cc5} gives us the identity
\begin{equation}
 (\pxm^3\gp)\{\ppp\gp,\gp\} \sim \gp[\gpp,\pxm^3\gpp].
\end{equation}
Now, with the help of this identity, we combine the last term in \eqref{cc4} with \eqref{cc6}, 
\begin{multline}\label{cb6}
 2i \gp[\pxm\gpp,\pxm^2\gpp] - \tfrac{2i}{3} \gp[\gpp,\pxm^3\gpp] 
 \\
 = -\tfrac{1}{3} \gp \left[ \pxu{\alpha} \pxu{\beta} \pxu{\gamma} \gpp , \pxd{\alpha} \pxd{\beta} \pxd{\gamma} \gpp \right],
\end{multline}
which   also   leads to a $SIM(1)$ invariant contribution to the action.
This completes our work because the action \eqref{caction} is a sum of 
\eqref{cb1a}, \eqref{cb1b}, \eqref{cb1c}, \eqref{cb2}, \eqref{cb3}, \eqref{cb4}, \eqref{cb5} and \eqref{cb6}.
The Chern-Simons action can now be written as 
\begin{multline}\label{caction}
 S_{CS} = \frac{k}{4\pi} \text{tr} \int\di^3x \pp \bigg(
  -2 \gxx \left( \pp \gxp \right)
  - \left(\pxx\gp\right)\gxp + \gp\left(\pxx\gxp\right)
\\
  - \frac{2}{3} \left( \pxx \gp \right) \left( \pxu{\alpha} \left\{ \gp , \pxd{\alpha} \gp \right\} \right)
  + 2i \gp \left[ \gxp, \gxx \right]
\\
  -2 \gp \left[ \pxu{\alpha} \gxp , \pxd{\alpha} \gxp \right]
  +2 \gp \left[ \pxu{\alpha} \gpp , \pxd{\alpha} \gxx \right]
\\
  + 2i \gp \left[ \pxu{\alpha} \pxu{\beta} \gpp , \pxd{\alpha} \pxd{\beta} \gxp \right]
  - \frac{1}{3} \gp \left[ \pxu{\alpha} \pxu{\beta} \pxu{\gamma} \gpp , \pxd{\alpha} \pxd{\beta} \pxd{\gamma} \gpp \right]
 \bigg).
\end{multline}
Thus, we have been able to write the Chern-Simons action explicitly using redefined $SIM(1)$ superfields. As we have expressed the Chern-Simons theory using redefined $SIM(1)$ superfields, it is manifestly $SIM(1)$  invariant.

\section{Conclusion} 
In this paper, we analysed   the Chern-Simons  theory in $SIM(1)$ superspace. We started with the  Lorentz invariant Chern-Simons  theory with      $\mathcal{N} =1$ supersymmetry. 
We broke the Lorentz symmetry down to the $SIM(1)$ symmetry,  and this in turn broke half the supersymmetry of the original theory. 
  Thus, we obtained a Chern-Simons theory with $\mathcal{N} =1/2$ supersymmetry. 
  This was the first time that a Chern-Simons theory with $\mathcal{N} =1/2$ supersymmetry has been constructed on a manifold 
  without a boundary. Theories with $\mathcal{N} =1/2$ supersymmetry are usually constructed by imposing 
  non-anticommutativity. 
 A non-anticommutative deformation of a four dimensional theory with $\mathcal{N} =1$ supersymmetry breaks down half of its supersymmetry, and we get a theory with $\mathcal{N} =1/2$ supersymmetry. 
  However, there are not enough degrees of freedom in three dimensions to perform such a deformation. 
  Thus, it is not possible to construct a Chern-Simons theory with $\mathcal{N} =1/2$ supersymmetry by imposing non-anticommutativity.
  It may be noted that we initially expressed the Chern-Simons theory using $SIM(1)$ projections of superfields. However, the transformation properties of these $SIM(1)$ projections were very complicated. So, we redefined these superfields to the ones which have simple $SIM(1)$ transformation properties. Finally, we expressed the Chern-Simons theory using these redefined $SIM(1)$ superfields. 
 
 It may be noted that the gauge sector of both the BLG theory \cite{1}-\cite{5}, 
 and the ABJM theory \cite{apjm}-\cite{abjm1} comprises of a Chern-Simons theory. 
 Furthermore, it is known that 
in string theory spontaneous breaking of the Lorentz symmetry  occurs due to an unstable perturbative string vacuum \cite{1a}. 
It is expected that such a mechanism can operate even in M-theory.
It has been demonstrated that  appropriate fluxes  can break the Lorentz symmetry  in M-theory \cite{mtheory}. 
The  spontaneous breaking of the Lorentz symmetry
can also be achieved by using a gravitational version of the Higgs mechanism for the 
 low energy effective  action of string theory \cite{2a}.  
A similar procedure can be followed for the eleven dimensional supergravity action, which is the 
 low energy effective  action of  M-theory.  Thus, it could be possible to study spontaneous symmetry breaking the
 Lorentz symmetry in  for M-theory. It would be interesting to use the results of this paper to study such a  
spontaneous symmetry breaking. It may be noted that another way to break the Lorentz symmetry would be to couple the BLG theory to a mass term, in such a 
 way that only $SIM(1)$ invariance is left. The mass deformed BLG theory is thought to be  related to the  theory of 
 multiple M5-branes \cite{blgmass}, hence, it would be interesting to study such a system. 
It could also be interesting to generalize the analysis of this paper to Chern-Simons theories with higher amount of supersymmetry. 
In fact,  $SIM(2)$ superspace has already been constructed  \cite{sup}, and from a supersymmetric point of view $\mathcal{N} =1$ supersymmetry 
in four dimensions is equivalent to $\mathcal{N} =2$ supersymmetry in three dimensions. So, it will be possible to use the results obtained in 
the construction of $SIM(2)$ superspace to analyse breaking of the Lorentz symmetry down to the 
$SIM(1)$ symmetry, for a Chern-Simons theory with $\mathcal{N} =2$ supersymmetry. It would be interesting to  use these results for analysing a system of multiple M2-branes. In this case, we can start by writing the ABJM theory or the BLG 
theory in $\mathcal{N} =2$ superspace, and then spontaneously  break the Lorentz symmetry to the $SIM(1)$ symmetry. It could also be interesting 
to investigate various mechanisms which can cause such a spontaneous  breaking of the Lorentz symmetry.

\end{document}